\documentclass[12pt,preprint]{aastex}
\slugcomment{Submitted to ApJ Letters: ~~\today}
\begin{document}

\title{GALACTIC CENTER EXTINCTION: EVIDENCE FOR METALLIC NEEDLES  IN THE GENERAL INTERSTELLAR MEDIUM }
%\author{Eli Dwek\altaffilmark{1} \& Joe Doe\altaffilmark{2}}
%\altaffiltext{1}{Code 685, Laboratory for Astronomy \& Solar Physics,
%NASA/Goddard Space Flight Center, Greenbelt, MD 20771, e--mail:
%eli.dwek@gsfc.nasa.gov}
%\altaffiltext{2}{SSAI, Code 685, NASA/Goddard Space Flight Center, 
%Greenbelt,
%MD 20771. e--mail: joedoe@stars.gsfc.nasa.gov}

\author{Eli Dwek}
\affil{Laboratory for Astronomy and Solar Physics, \\ NASA Goddard Space Flight Center,
Greenbelt, MD 20771 \\ e-mail: eli.dwek@nasa.gov}

\begin{abstract}
The extinction curve derived from {\it ISO} mid-infrared (IR) observations of the Galactic center (GC) exhibits a surprisingly flat behavior in the $\sim$ 3 to 8 $\mu$m region, contrary to the deep minimum expected from standard interstellar dust models consisting of bare silicate and graphite dust particles. We show that this extinction is likely caused by the presence of metallic needles in the interstellar medium (ISM) towards the Galactic center. If the needles contribute only to the 3 -- 8~$\mu$m extinction, they must have a long wavelength cutoff at $\sim$~8~$\mu$m, and therefore a typical length over radius ratio of $\sim$~600, smaller than the $\sim 3\times 10^3$ aspect ratio determined for the needles in Cas~A. Homogeneously distributed throughout the ISM, they comprise only a minor mass fraction of the ISM, with a needle-to-H mass ratio of $\sim$ ~5$\times~10^{-6}$, which is equivalent to 0.14\% of the silicate dust mass. Their total ISM abundance can then be readily explained by the combined production in SNe and O-rich stellar outflows. It is currently unclear how universal the GC extinction law is. Local 2 -- 5~$\mu$m extinction measurements seem to be consistent with the standard extinction law, suggesting a non uniform distribution of needles in the ISM.  The GC observations show that metallic needles, in spite of their low abundance or non uniform distribution, can be the dominant source of opacity in the 3 - 8~$\mu$m wavelength region. However, expelled into the intergalactic medium, their abundance is too low to cause any dimming of cosmological sources, and their length is too short to make them a significant source of submillimeter emission.   
\end{abstract}
\keywords{ISM: Galactic center -- dust, extinction -- infrared: ISM}

\section{INTRODUCTION}
Analysis of the $\sim$ 2.6 to 9 $\mu$m spectrum of the Galactic center (GC) obtained with the Short Wavelength Spectrometer (SWS) on board the {\it Infrared Space Observatory} ({\it ISO}) by Lutz et al. (1996) and Lutz (1999) has revealed an anomalous extinction law, compared with that expected from standard interstellar dust models consisting of a population of bare silicate and graphite grains and polycyclic aromatic hydrocarbons (PAHs) (Li \& Draine 2001; Zubko, Dwek, \& Arendt 2004, hereafter ZDA). 
The  GC extinction is characterized by a relatively flat behavior at $\sim$ 3 -- 9 $\mu$m, contrary to model predictions of a deep minimum in that wavelength region. The minimum results from the steep decline in the graphite and silicate extinction from UV-optical to near--IR wavelengths, which is followed by a sharp rise at $\sim$ 10~$\mu$m caused by the silicate 9.7~$\mu$m absorption feature (see Figure 1). 
ZDA have shown that an interstellar dust model consisting of bare silicate and amorphous carbon dust,  PAHs, and composite  particles containing silicates, organic refractories, water ice, and voids, can reproduce the GC extinction curve (see model COMP-AC-x in Figure 21 of ZDA). Here we pursue an alternative explanation, namely that the ``anomalous"  extinction is produced by a population of metallic needles in the interstellar medium (ISM).

%==== Figure 1 ==== 
\begin{figure}
\plotone{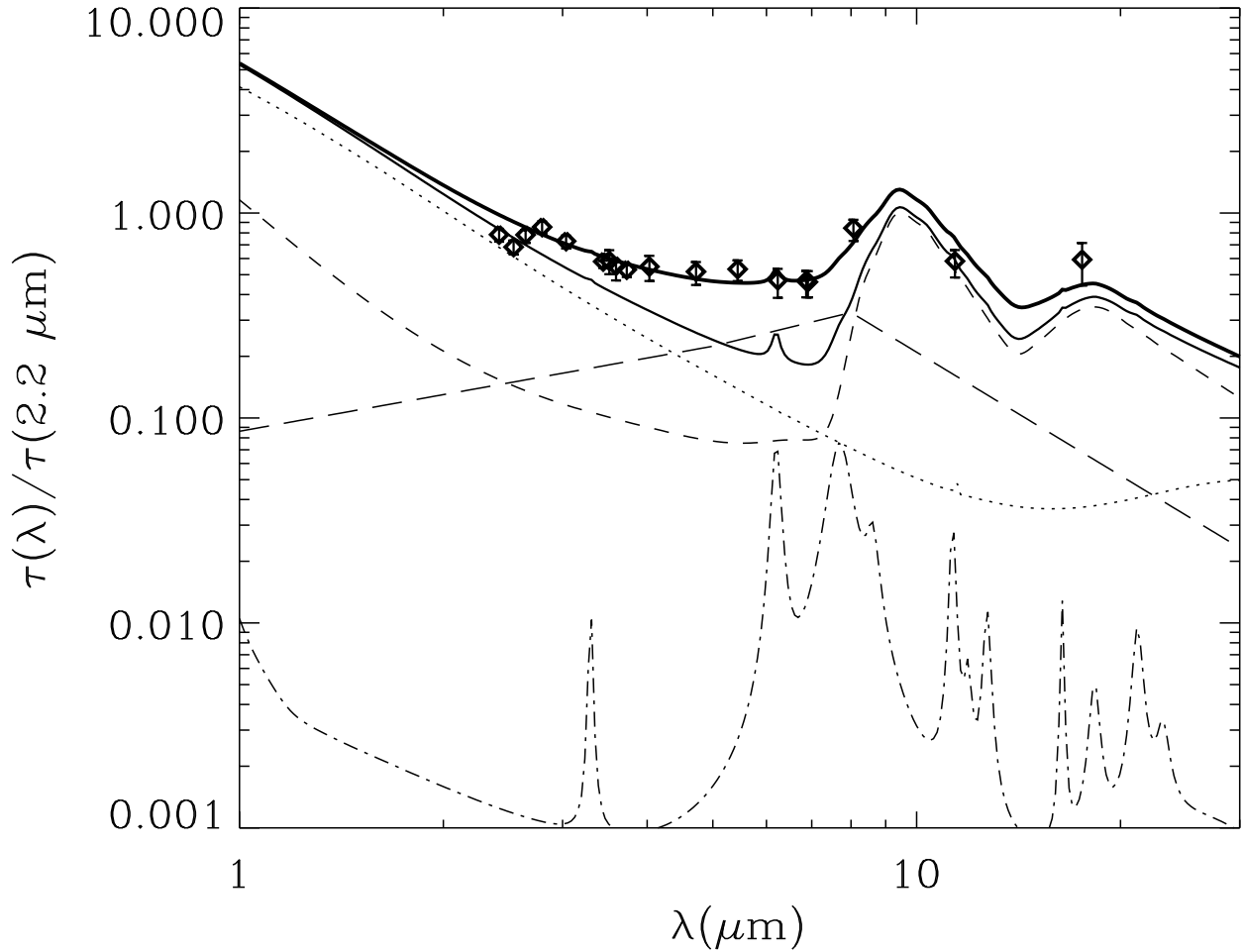}
\caption{Extinction predicted by the BARE-GR-S dust model of Zubko, Dwek, \& Arendt (2004; ZDA) with the additional population of metallic needles is compared to the Galactic center extinction derived by Lutz (1999). The various contributors to the extinction include PAHs (dashed-dotted line), graphite (dotted line), silicates (short dashed line), and metallic needles (long dashed line). The standard extinction predicted by the BARE-GR-S model is shown as a thin solid line, and falls short of the observations. The thick solid line shows is the total extinction obtained with the addition of needles to the population of interstellar dust particles.}
\end{figure}

 Metallic needles (whiskers) play an important role in Hoyle-Narlikar cosmologies. Their presence in the intergalactic medium has been invoked by Hoyle \& Wickramasinghe (1999) to create the cosmic microwave background (CMB) by the thermalization of starlight. Needles can, in principle, provide the needed opacity at high redshift (Wright 1982), however, Li (2003) recently raised questions about the applicability of the antenna theory to modeling the optical properties of exceedingly thin needles, and the amount of intergalactic iron needed to produce the CMB. Banerjee et al. (2000)  suggested that intergalactic absorbing needles can explain to the observed redshift-magnitude relation of Type Ia supernovae (Riess et al. 1998, Perlmutter et al. 1999) without the need to invoke a positive cosmological constant. More recently, Narlikar et al. (2003) suggested that needles can also explain the CMB anisotropy detected by the {\it Wilkinson Microwave Anisotropy Probe} (Bennett et al. 2002). 
 
Here we suggest a more modest role for needles as a minor dust constituent in the Galactic ISM that nevertheless may be the dominant source of extinction in the $\sim$ 3 to 8~$\mu$m wavelength region. Conductive metallic needles have recently been  proposed by Dwek (2004) as the source of the submillimeter emission  detected with the SCUBA (Dunne et al. 2003) from Cas~A. This, and the possible detection of needles in the ejecta of SN1987A (Wickramasinghe \& Wickramasinghe 1993) may suggest that needles can be readily created in supernovae. Furthermore, Kemper et al. (2002) have suggested that non-spherical metallic particles can provide the 3 - 8~$\mu$m opacity needed to fit the spectral energy distribution of the OH/IR star OH~127.8+0.0, suggesting that metallic needles may also be produced in quiescent O-rich stellar outflows. Conductive needles may therefore be a noticeable dust constituent of the general interstellar medium (ISM).

Needle properties required to explain the GC extinction curve are derived in \S2, and the astrophysical implication, whether the GC extinction is a special case or more characteristic of the general ISM is discussed in \S3.   
  
\section{NEEDLE PROPERTIES}

The extinction optical depth $\tau(\lambda)$ at wavelength $\lambda$ to a source located at distance $S$ due to an assembly of dust particles along the line of sight (LOS) is given by:
%---------------------
\begin{equation}
\tau(\lambda) = \sum_j \int_0^L \ n_H(s) {\rm d}s\ \left[\int_{a_{min,j}}^{a_{max,j}} {\rm d}a\ f_j(a)\ \kappa_j(\lambda,\ a)\  m_{d,j}(a,\ s) \right]
\end{equation}
%---------------------
where the sum is over dust composition $j$. The integral in square brackets is over the grain size distribution $f_j(a)$ (which is normalized to unity over the \{$a_{min,j}$, $a_{max,j}$\} size interval), $m_{d,j}(a,\ s) = 4\pi \rho_j a^3/3$ is the mass of a dust particle of radius $a$ and mass density $\rho_j$ located at $s$, and $\kappa(\lambda,\ a)$ is the mass extinction coefficient of the dust. The outer integral is over the LOS, where $n_H(s)$ is the H-number density at distance $s$.

Equation (1) can be written in a simplified form as:
%--- equation 2 ---
\begin{eqnarray}
\tau(\lambda) & = & \sum_j \kappa_j(\lambda){\cal M}_{d,j}(LOS) \\ \nonumber
 & = &  \sum_j \kappa_j(\lambda) \rho_{d,j} L
\end{eqnarray}
where  $\kappa_j(\lambda)$ represents an average over the grain size distribution, and ${\cal M}_{d,j}$ is the mass column density of the dust of composition $j$,   $\rho_{d,j}$ is the average dust mass density in the ISM, and $L$ is the total distance along the LOS.

Figure 1 depicts the average interstellar extinction curve for a mixture of bare graphite and silicate particles and PAHs that satisfies the UV-optical extinction curve, the diffuse IR emission, as well as the solar interstellar abundances constraints (model BARE-GR-S in Zubko, Dwek, \& Arendt 2004, ZDA). The extinction was normalized to unity at K ($\lambda= 2.2\ \mu$m), for which $\tau$(K)=3.47/1.086=3.20 (Rieke, Rieke, \& Paul 1989, Lutz et al. 1996). At $\sim10\ \mu$m the extinction is dominated by the silicate absorption feature which has a mass absorption coefficient $\kappa_{sil}(9.7\ \mu$m) = 3$\times10^3$ cm$^2$ g$^{-1}$. Also shown in the figure are the new improved extinction measurements derived from the SWS recombination line data presented by Lutz (1999). These measurements reinforce the previous results (Lutz et al. 1996) that the $\sim$ 3 - 8~$\mu$m wavelength extinction is essentially flat, with an average value of $\sim 0.6\times\ \tau$(K), and lacks the deep $\sim$ 7~$\mu$m minimum characterizing the standard bare graphite-silicate interstellar dust model. 

Metallic needles with appropriate physical parameters can fill in the minimum in the ``standard" interstellar extinction curve and produce the observed extinction towards the GC. At wavelengths below the long wavelength cutoff $\lambda_0$ (see below) one can use Mie calculations for infinitely long cylinders to calculate their optical properties. Their extinction is dominated by absorption, and for thin needles with a resistivity $\rho_R = 1\times10^{-18}$, and length $\ell >> a$, where the radius $a$ = 0.01 $\mu$m,  the Mie calculations give a mass absorption coefficient that slowly rises from values of $\sim  2\times10^5$, $\sim 5 \times10^5$, and $\sim 9 \times10^5$~cm$^2$~g$^{-1}$  at 2, 5 and 10~$\mu$m, respectively, to a value of  $\sim 3 \times10^6$~cm$^2$~g$^{-1}$  at 30~$\mu$m, (Hoyle \& Wickramasinghe 1999). 

The GC extinction measurement provide important constraints of the needle properties. The Cas~A needle were required to be efficient submillimeter emitters, with a long wavelength cutoff in their mass absorption coefficient at 400~$\mu$m (Dwek 2004). Figure 1 shows that the derived $\lambda > 8.0~\mu$m extinction is in fairly good agreement with that predicted by  the standard bare silicate-graphite model. If the near-IR extinction were to be fitted with Cas~A type needles, then their  rising extinction efficiency at longer wavelengths would produce a $\sim$ 20~$\mu$m extinction that exceeds the observed value by a factor of three. Since the needles cannot contribute significantly to the opacity at mid-IR wavelengths, the cutoff wavelength $\lambda_0$ should be less than about 8~$\mu$m. In fact, an excellent fit to the GC extinction is obtained with $\lambda_0 = 8\ \mu$m, after which the absorptivity of needles drops off as $\lambda^{-2}$.
The cutoff wavelength, $\lambda_0$, is given by:
%--- equation 4 ---
\begin{equation}
\lambda_0 = {1\over 2}\ \rho_Rc\ {(\ell/a)^2 \over \ln(\ell/a)} \qquad ,
\end{equation}
giving a value of $\ell/a$~=~580 for a typical needle resistivity of $1\times10^{-6}\ \Omega$ cm $\approx 1 \times 10^{-18}$ s. This value is smaller than the value of $\ell/a$~=~3000 found for the needles in Cas~A, suggesting that grain-grain collisions or other break up mechanisms play an important role in the evolution of needles in the ISM. 

Figure 1 shows the contribution of the various dust components to the Galactic center extinction. The total extinction without the needles (thin solid line) exhibits a minimum at $\sim$ 7~$\mu$m. This minimum is filled by the opacity of the needles (long dashed curve). The total extinction which includes the population of needles is shown as a thick solid line in the figure.

\section{DISCUSSION}
An important quantity is the abundance of the needles needed to give rise to the observed GC extinction. Given a lifetime for their destruction in the general ISM, their abundance can be used to estimate their required production rate in supernovae and in stellar outflows. Furthermore, if intergalactic needles originate from galactic outflows, their abundance in the ISM of galaxies can be used to set an upper limit on their abundance in the general intergalactic medium. This upper limit can, in turn, be used to set an upper limit on the intergalactic opacity that can be attributed to metallic needles.

Using eq. (2) we can write the needles-to-silicate mass ratio as:
\begin{equation}
{\rho_{ndl}\over \rho_{sil} }= \left[{\tau_{ndl}(\lambda) \over \tau_{sil}(\lambda)}\right]\times \left[{\kappa_{sil}(\lambda) \over \kappa_{ndl}(\lambda)}\right]\times \left[{L_{sil} \over L_{ndl}}\right]
\end{equation}
At $\lambda$ = 9.7~$\mu$m, $\kappa_{sil}$ = 3$\times10^3$ cm$^2$~g$^{-1}$, and $\kappa_{ndl}$ = 5.0$\times10^5$ cm$^2$~g$^{-1}$. Figure 1 shows that for needles $\tau_{ndl}(9.7\ \mu$m) = 0.22$\times$3.20 = 0.70, whereas for the silicates  $\tau_{sil}(9.7\ \mu$m) = 3.1. 

If we assume that the needles are uniformly mixed with the silicate dust, so that $L_{ndl} = L_{sil}$, we get that needles contribute only a minor fraction to the total mass of ISM dust which, normalized to the abundance of silicates, is given by:
\begin{equation}
{\rho_{ndl}\over \rho_{sil} }= 1.4\times 10^{-3} \qquad \qquad {\rm for\ needles\ uniformly\ mixed\ in\ the\ ISM}
\end{equation}
The silicate-to-H mass ratio is 0.004 (model BARE-GR-S of ZDA), giving a needle-to-H mass ratio of 5.6$\times$10$^{-6}$.

Alternatively, we can assume that the abundance of needles is comparable to that of the silicates,  i.e. $\rho_{ndl} \approx \rho_{sil}$, in which case equation (4) implies that the needles are inhomogeneously distributed in the general ISM, and occupy a small fraction of the LOS to the Galactic center, given by:
\begin{equation}
{L_{ndl}\over L_{sil} } \approx 1.4\times 10^{-3}  \qquad {\rm for\ a\ patchy\ distribution\ of\ needles\ with} \ \ \rho_{ndl} \approx \rho_{sil}
\end{equation}

The universality of the Lutz extinction law towards the Galactic center is still an unresolved issue. Analysis of stellar data (Whittet 1988) show that the 1.6 to 5~$\mu$m extinction towards stars within a distance of 3~kpc is fairly uniform, and closely represented by a $\lambda^{-1.7}$ extinction law, similar to the standard extinction curve in Fig. 1. A similar extinction law was derived by Rieke, Rieke, \& Paul (1989) in the same wavelength region from observations of  select GC sources. Extended to 8~$\mu$m, these observations will argue against the universality of the GC extinction law. However, a major uncertainty in these extinction measurements is the intrinsic spectrum of the background stars, specifically the amount of excess photospheric IR emission, which may be variable.  Near-IR (3 -- 8~$\mu$m) extinction derived from the measurements of line ratios may therefore be more reliable than those derived from stellar continua, an issue that can only be resolved by more observations of background sources with well understood stellar spectrum.  After the initial submission of this paper for publication, Indebetouw et al. (2004) presented the 1.25 to 8.0 $\mu$m extinction derived from data obtained by the {\it Spitzer} along the two $\ell$ = 42$\arcdeg$ and 284$\arcdeg$ lines of sight in the Galactic plane. The two extinction laws were found to be consistent with the Lutz (1999) measurements towards the GC, strengthening the case for the universality of this law in our Galaxy. 

Whether the needles are uniformly or inhomogeneously distributed in the ISM, the upper limit to the total mass of needles in the Galaxy is about  3$\times$10$^4$~M$_{\odot}$, assuming an ISM mass of 5$\times$10$^9$~M$_{\odot}$ (Sodroski et al. 1997). The average lifetime of interstellar dust is about (4--6)$\times$10$^{8}$~yr (Dwek 1998, Jones 2001), requiring a needle production rate of $\sim$6$\times$10$^{-4}$~M$_{\odot}$~yr$^{-1}$. The SCUBA observations of Cas~A suggest that about 10$^{-3}$~M$_{\odot}$ of metallic needles formed in the SN ejecta. If the Cas~A yield is typical, and if a comparable production rate takes also place in quiescent O-rich stellar outflows, then a Galactic SN rate of about 0.03~yr$^{-1}$ will be sufficient to account for the mass of ISM needles. 

It is interesting to examine whether the needles can provide any significant extinction to cosmological sources.
The optical depth to redshift $z$ is given by:
\begin{eqnarray}
\tau(\nu_0,\ z) & = & \int_0^z \sigma[\nu_0/(1+z')]\ n_n(z')\  c\ \left| {{\rm d}t\over {\rm d}z}\right| \\ \nonumber
  & = & \kappa_n(\nu_0)\ \rho_n\ R_H \int_0^z {(1+z')^2\over E(z') }\ {\rm d}z'
\end{eqnarray}
where $\sigma(\nu)$ is the needles' cross section [assumed to be flat in the $\nu_0$ to  $\nu_0/(1+z)$ frequency range],  $n_n(z) \propto (1+z)^3$ is their number density in the intergalactic medium, $\kappa_n$ is their mass absorption coefficient, $\rho_n$ is their present mass density, and $R_H\equiv c/H_0$ is the Hubble radius. The function $E(z) = [(1+z)^2(\Omega_m z+1) - z (2+z)\Omega_{\Lambda}]^{1/2}$, where $\Omega_m$ and $\Omega_{\Lambda}$ are, respectively, the matter density and the the cosmological constant, normalized to their critical values. Using standard $\Lambda$CDM cosmology with $\Omega_m$ = 0.27 and $\Omega_{\Lambda}$ = 0.73, the optical depth to a hypothetical source at redshift $z$ = 2 is given by $\tau = 4.5\ \kappa_n \rho_nR_H$. 
The dust-to-gas mass ratio of needles in the Milky Way ISM sets an upper limit on their abundance in the  intergalactic medium. The baryonic mass density for $H_0$ = 70 km s$^{-1}$ Mpc$^{-1}$ is 4.2$\times 10^{-31}$ g~cm$^{-3}$, giving an intergalactic needle mass density of  $\rho_n$ = 2.9$\times 10^{-36}$~g~cm$^{-3}$.  The Hubble radius $R_H$ = 4290~Mpc, giving a mass column density of $\sim$ 2$\times 10^{-7}$~g~cm$^{-2}$ out to $z$ = 2. For a value of $\kappa$(V) $\approx 10^5$ cm$^2$~g$^{-1}$, the visual optical depth to that distance will be about 0.02. The needles giving rise to the GC extinction have therefore a negligible effect on the dimming of light from cosmological sources, in particular high-$z$ Type~Ia supernovae. They are also relatively short compared to the needles in the Cas~A ejecta, and will therefore not give rise to any significant submillimeter emission.

In summary, the anomalously flat 3 -- 8~$\mu$m  extinction observed by Lutz et al. (1996) and Lutz (1999) towards the Galactic center and by Indebetouw et al. (2004) towards the $\ell$ = 42 $\arcdeg$ and 284$\arcdeg$ lines of sight in the Galactic plane can be attributed to the presence of relatively short metallic needles in the ISM. The universality of this extinction law, and consequently the distribution of these needles in the general ISM is still a subject of ongoing investigation, that could be resolved with further {\it Spitzer} observations.

Acknowledgement: I thank Rick Arendt for useful discussions and Ciska Kemper for bringing to my attention the contribution of metallic particles to the 3 -- 8~$\mu$m opacity in OH~127.8+0.0. I also thank Dieter Lutz for helpful correspondence, and for providing the extinction data in digital format. ED acknowledges NASA's Long Term Space Astrophysics (LTSA) Program NRA 2004-OSS-01 for support of this work.

\end{document}